\documentclass[fleqn,twoside]{article}
\usepackage{espcrc2}
\usepackage{latexsym}
\usepackage{graphicx}

\usepackage{graphicx}
\usepackage[figuresright]{rotating}

\newcommand{\AmS}{{\protect\the\textfont2
  A\kern-.1667em\lower.5ex\hbox{M}\kern-.125emS}}

\newcommand{\Dslash}{\rlap{/}\kern-2.0pt D}

\def\mres{m_{\rm res}}
\def\GeV{\rm GeV}
\def\MeV{\rm MeV}
\def\LorInd{{\mu_1\mu_2\cdot\cdot\cdot\mu_n}}
\def\LorIndtwo{{\mu_2\cdot\cdot\cdot\mu_n}}
\def\Dcc{\stackrel{\,\leftrightarrow}{D}}

\title{Nucleon matrix elements with domain wall fermions}

\author{Konstantinos Orginos\address[RBRC]{RIKEN-BNL Research Center, Brookhaven National Laboratory, Upton, NY 11973, USA} %
for the RBC collaboration\thanks{The current members of the RBC collaboration are: Y.~Aoki, T.~Blum, N.~Christ, M.~Creutz, C.~Dawson, T.~Izubuchi, L.~Levkova, X.~Liao, G.~Liu, R.~Mawhinney, Y.~Nemoto, J.~Noaki, S.~Ohta, K.~Orginos, S. Prelovsek, S.~Sasaki and A.~Soni. We thank RIKEN, BNL and the U.S. DOE for providing the facilities essential for the completion of this work.}}
       
\begin{document}

\begin{abstract}
We present the status of our calculation of the first few
moments of the nucleon structure functions. Our calculations
are done using domain wall fermions in the quenched approximation
with the DBW2 gauge action at 1.3GeV inverse lattice spacing.
\vspace{-1pc}
\end{abstract}

\maketitle

The structure of the nucleon is one of the fundamental problems
that lattice QCD can address. In the last few years, substantial efforts
have been made by several groups~\cite{Horsley:Lat2002,Gockeler:2000ja,Gockeler:1996wg,Dolgov:2002zm} in calculating the non-perturbative
matrix elements relevant to nucleon structure.
Up to now only Wilson fermions, improved and unimproved, have been
used in both the quenched approximation and in full QCD.
In this report we examine the feasibility of studying nucleon
matrix elements with domain wall fermions in the quenched approximation.
Domain wall fermions have only ${\cal O}(a^2)$ lattice artifacts,
non-perturbative renormalization works very well, and have
no problem with exceptional configurations~\cite{Blum:2000kn,Blum:2001sr}. 
Furthermore, the chiral symmetry
they preserve on the lattice eliminates mixings with lower
dimensional operators, rendering the renormalization of certain
matrix elements significantly simpler. For the above reasons,
a study of the nucleon structure with domain wall fermions is
definitely important.  

We study the nucleon matrix elements relevant to the leading
twist contributions to the moments of the nucleon structure functions.
The leading twist matrix elements are:
\begin{eqnarray}
 && \frac{1}{2} \sum_s \langle p,s|{{\cal O}^{q}_{\{\LorInd\}}}
 |p,s\rangle=
 2 {\langle x^{n-1}\rangle_q}(\mu)\times\nonumber\\&&\times
 [ p_{\mu_1}p_{\mu_2}
\cdot\cdot\cdot p_{\mu_n}+
\cdot\cdot\cdot -tr]\nonumber\\
 &&-\langle p,s|{{\cal O}^{5q}_{\{\sigma\LorInd\}}} |p,s\rangle =
 \frac{2}{n+1}{\langle x^{n}\rangle_{\Delta q}}(\mu)\times\nonumber\\&&\times
 [ s_\sigma p_{\mu_1}p_{\mu_2}\cdot\cdot\cdot p_{\mu_n}+\cdot\cdot\cdot
 -tr]\nonumber\\
 &&\langle p,s|{\cal O}^{[5]q}_{[\sigma\{\mu_1]\LorIndtwo\}}
 |p,s\rangle=
 \frac{1}{n+1}{d_n^{ q}}(\mu)\times\nonumber\\&&\times
 [ (s_\sigma p_{\mu_1} - s_{\mu_1} p_{\sigma})p_{\mu_2}
\cdot\cdot\cdot p_{\mu_n}+\cdot\cdot\cdot -tr]\nonumber\\
&&\langle p,s|{{\cal O}^{\sigma q}_{\rho\nu\{\LorInd\}}}
 |p,s\rangle =
 \frac{2}{m_N}{\langle x^{n}\rangle_{\delta q}}(\mu)\times\nonumber\\&&\times
 [(s_\rho p_{\nu} - s_{\nu} p_\rho)p_{\mu_1}p_{\mu_2}\cdot\cdot\cdot p_{\mu_n}
+\cdot\cdot\cdot -tr]\nonumber
\end{eqnarray}
where $p_\mu$ and $s_\mu$ are the nucleon momentum and spin vectors,
$m_N$ the nucleon mass, and
\begin{eqnarray}
  {\cal O}^q_\LorInd\!\!\!\!\!\!  &=&\!\!\!\!\!\!
 \left(\frac{i}{2}\right)^{n-1}\bar{q}\gamma_{\mu_1} 
\Dcc_{\mu_2}\cdot\cdot\cdot \Dcc_{\mu_n}q -tr\nonumber\\
{\cal O}^{5q}_{\sigma\LorInd} \!\!\!\!\!\! &=&\!\!\!\!\!\!
 \left(\frac{i}{2}\right)^{n}\bar{q}\gamma_{\sigma}\gamma_5 
 \Dcc_{\mu_2}\cdot\cdot\cdot \Dcc_{\mu_n}q -tr
\nonumber\\
{{\cal O}^{\sigma q}_{\rho\nu\LorInd}}\!\!\!\!\!\! &=&\!\!\!\!\!\!
\left(\frac{i}{2}\right)^{n}\bar{q}\gamma_5 \sigma_{\rho\nu} \Dcc_{\mu_1}\cdot\cdot\cdot \Dcc_{\mu_n}q - tr\nonumber
\end{eqnarray}
$\{\}$ implies symmetrization and $[]$ implies anti-symmetrization.
For the conventions used see ~\cite{Dolgov:2002zm}.

Our current results are restricted only to those matrix elements that
can be computed with zero momentum nucleon states. We use the DBW2
gauge action which is known to improve the domain wall fermion chiral
properties~\cite{Orginos:2001xa,Aoki:2001dx}. These results are from 204 
lattices of size 
$16^3\times 32$ at $\beta=0.870$ with lattice spacing
$a^{-1}=1.3\GeV$.  This provides us with a physical volume ($\sim
(2.4fm)^3$) large enough to reduce finite size effects known to affect
some nucleon matrix elements such as
$g_A$~\cite{Sasaki:2001th,Ohta:Lat2002}.  Using fifth dimension length
$L_s=16$ we achieve a residual mass $\mres \sim
0.8\MeV$~\cite{Orginos:2001xa,Aoki:2001dx}.  The quark masses used are
$m_q a = 0.02, 0.04, 0.06, 0.08,$ and $0.10$ which give pion masses
ranging from $390\MeV$ to $850\MeV$.  We used Coulomb gauge fixed box
sources with size $8^3$ which have been shown to couple very well to
the nucleon ground state~\cite{Aoki:2001dx}.  The source - sink
separation was set to 10 time slices or $\sim 1.5fm$ in physical
units. Finally in order to construct the three point functions we used
sequential propagators~\cite{Bernard:1985ss,Martinelli:1989rr} with
point sinks.  With theses choices the signal to noise ratio for the
three point functions in the plateau region is about 10 for the
lightest quark mass.  A part of our code tests was a small quenched
run with Wilson fermions and Wilson gauge action at $\beta=6.0$. We
were able to reproduce the results
in~\cite{Gockeler:1996wg,Dolgov:2002zm}.

\begin{figure}[t]
\includegraphics[width=\columnwidth]{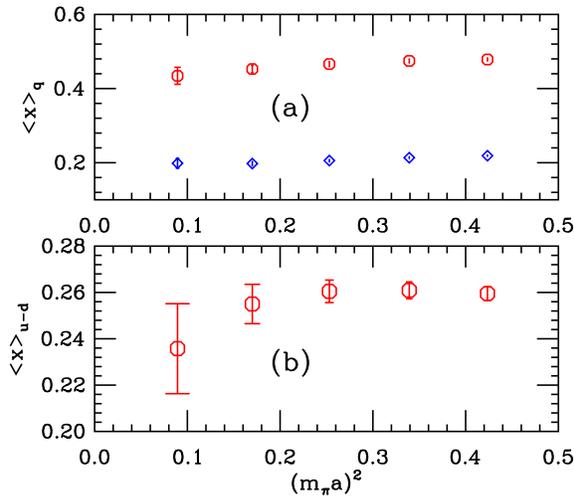}
\vspace{-30pt}
\caption{Quark density $\langle x \rangle$ vs. the pion mass squared. 
(a) The connected up (octagons) and down (diamonds) quark contributions. 
(b) The flavor non-singlet $\langle x \rangle_{u-d}$.}
\label{fig:X_ns_mpi2}
\vspace{-20pt}
\end{figure}

Fig.~\ref{fig:X_ns_mpi2} presents our results for the quark
density distribution $\langle x \rangle_q$. This is related to the
lowest moment of the unpolarized structure funtions $F_1$ and
$F_2$. We plot the unrenormalized result for $\langle x\rangle_u$, 
$\langle x \rangle_d$ and the flavor non-singlet $\langle
x \rangle_{u-d}$.  The latter exibits a noticable curvature for the
two lighter quark masses indicating that the quenched result for
$\langle x \rangle_{u-d}$ may be closer to the phenomenological
expectations than previously
thought~\cite{Gockeler:1996wg,Dolgov:2002zm}. This curvature may be
the first indication of the chiral log behaviour that has to set in at
sufficient small quark
masses~\cite{Detmold:2001jb,Detmold:2002nf,Thomas:2002sj,Schierholz:Lat2002}.
The ratio
$\frac{\langle x \rangle_u}{\langle x \rangle_d}$ is $2.37(6)$ at the
chiral limit in agreement the quenched Wilson fermion
result~\cite{Gockeler:1996wg,Dolgov:2002zm}.

%\begin{figure}[t]
%\vspace{9pt}
%\includegraphics[width=\columnwidth]{1Dq_vs_mpi2.ps}
%\caption{Helicity zero moment}
%\label{fig:1Dq_vs_mpi2}
%\end{figure}
\begin{figure}[t]
\includegraphics[width=\columnwidth]{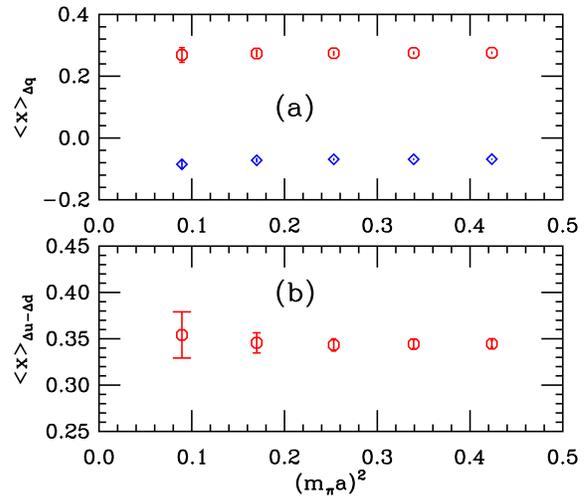}
\vspace{-30pt}
\caption{Helicity  $\langle x \rangle_{\Delta q}$ vs. the pion mass squared. 
(a) The connected up (octagons) and down (diamonds) quark 
contributions. (b) The flavor non-singlet 
$\langle x \rangle_{\Delta u- \Delta d}$. }
\label{fig:XDq_vs_mpi2}
\vspace{-20pt}
\end{figure}

We measure also the helicity distributions $\langle 1 \rangle_{\Delta q}$
and   $\langle x \rangle_{\Delta q}$. 
%Fig.~\ref{fig:1Dq_vs_mpi2} shows
%the renormalized results for $\langle 1 \rangle_{\Delta q}$. 
A detailed discussion of our results for  $\langle 1 \rangle_{\Delta q}$ 
can be found in~\cite{Ohta:Lat2002}.
In Fig.~\ref{fig:XDq_vs_mpi2} we present our unrenormalized data
for $\langle x \rangle_{\Delta q}$. Unlike the quark density distributions
we do not see a significant dependence on the quark mass. On the basis of 
chiral perturbation theory arguments this matrix element is indeed expected
to show the chiral log behavior at smaller quark masses than the 
quark density distributions~\cite{Detmold:2002nf,Thomas:2002sj}. 
The ratio 
$\frac{\langle x \rangle_{\Delta u}}{\langle x \rangle_{\Delta d}}$ 
is roughly $-4$, consistent with other lattice 
results~\cite{Gockeler:2000ja,Dolgov:2002zm}.

\begin{figure}[t]
\includegraphics[width=\columnwidth]{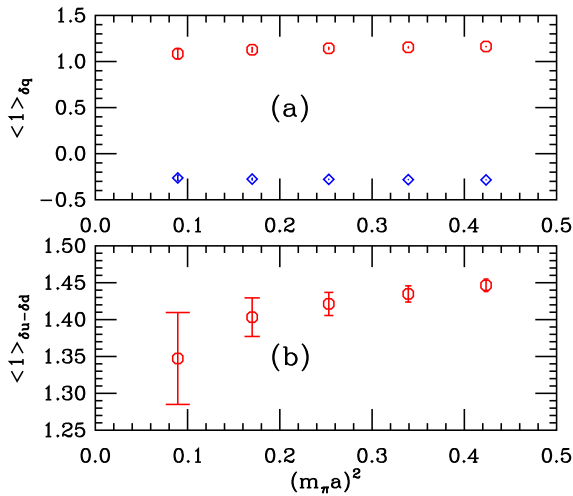}
\vspace{-30pt}
\caption{Transversity  $\langle 1 \rangle_{\delta q}$ vs. 
the pion mass squared. (a) The connected up (octagons) 
and down (diamonds) quark contributions. (b) The flavor non-singlet 
$\langle 1 \rangle_{\delta u- \delta d}$.}
\label{fig:1dq_vs_mpi2}
\vspace{-22pt}
\end{figure}

The lowest moment of the transversity $\langle 1 \rangle_{\delta q}$
is also measured. In Fig.~\ref{fig:1dq_vs_mpi2} we plot the
unrenormalized contributions for both the up and down quark, and the
flavor non-singlet combination 
$\langle 1 \rangle_{\delta u - \delta d}$. 
Again the quark mass dependence is very mild and there is no sign of
a chiral log behavior. The ratio 
$\frac{\langle 1 \rangle_{\delta u}}{\langle 1 \rangle_{\delta d}}$ 
is also roughly $-4$.

\begin{figure}[t]
\includegraphics[width=\columnwidth]{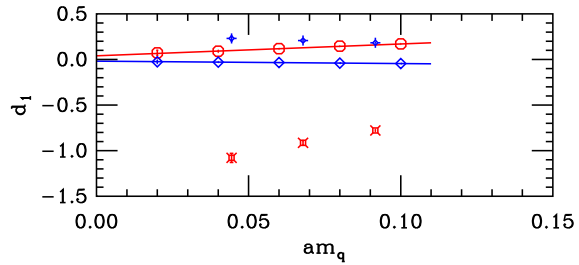}
\vspace{-35pt}
\caption{The connected $d_1$ matrix element vs. quark mass for the  up (octagons) and down (diamonds) quarks. The up (fancy squares) and down (fancy diamonds) quark for Wilson fermions~\cite{Dolgov:2002zm}.}
\label{fig:d1}
\vspace{-22pt}
\end{figure}

Finally we computed the $d_1$ matrix element which is a twist 3 contribution
to the first moment of $g_2$. If chiral symmetry is broken the
operator 
$$
 {\cal O}^{[5]q}_{34} = \frac{1}{4}\bar{q}\gamma_5\left[\gamma_3\Dcc_4-\gamma_4\Dcc_3\right]q
$$
which is used to measure $d_1$ mixes with the lower dimensional operator 
$
 {\cal O}^{\sigma q}_{34} = \bar{q}\gamma_5\sigma_{34}q.
$ 
Hence in  Wilson fermion calculations a non perturbative
subtraction has to be
performed~\cite{Gockeler:2000ja,Horsley:Lat2002}.  With domain wall
fermions this kind of mixing is proportional to the residual mass
which in our case is negligible. Thus, we expect that a
straight forward computation of $d_1$ with domain wall fermions
provides directly the physically interesting result. In Fig.~\ref{fig:d1}
we present our unrenormalized results for $d_1$ 
as a function of the quark mass.  For
comparison we also plot the unsubtracted quenched Wilson results for
$\beta=6.0$ from~\cite{Dolgov:2002zm}. The fact that our result almost
vanishes at the chiral limit is an indication that the power
divergent mixing is absent for domain wall fermions.
The behavior we find for the $d_1$ matrix element is consistent
with that of the subtracted $d_2$ measured 
by QCDSF~\cite{Gockeler:2000ja,Horsley:Lat2002} with Wilson fermions.

In conclusion, we have started the computation of moments of nucleon
structure functions with domain wall fermions. Our current results are
unrenormalized and restricted to those matrix elements that can be
computed with zero momentum nucleon states. Yet we already have hints
of a couple of potentially interesting results. First we have an
indication of the possible onset of chiral log behavior for $\langle
x\rangle_{u-d}$.  Also it is very encouraging, although expected, to
see the lack of power divergent contributions for $d_1$. Our project
is ongoing. We hope to have more statistics and non-perturbative
renormalization of the presented matrix elements in the near future.

\end{document}